\documentclass[preprint,review,11pt,3p]{elsarticle}
\usepackage{amsmath,amssymb,amsfonts}
\usepackage{graphicx}
\usepackage{textcomp}
\usepackage{algorithmic}
\usepackage{algorithm}
\usepackage[table]{xcolor}
\usepackage{color,soul}
\usepackage{multirow}
\usepackage{tabularx}
\usepackage{booktabs}
\usepackage{array}

\usepackage{tikz}

\begin{document}

\begin{frontmatter}

\title{Topology-Aware Loss for Segmentation in Computed Tomography Images}

\author[inst1]{Seher Ozcelik}
\ead{sozcelik19@ku.edu.tr}
\author[inst2]{Sinan Unver}
\ead{sunver@ku.edu.tr}
\author[inst3]{Ilke Ali Gurses}
\ead{igurses@ku.edu.tr; iagurses@gmail.com}
\author[inst4]{Rustu Turkay}
\ead{rustu.turkay@sbu.gov.tr; rustuturkay@hotmail.com}
\author[inst5]{Cigdem Gunduz-Demir\corref{cor1}}
\ead{cgunduz@ku.edu.tr}

\affiliation[inst1]{organization={Computational Sciences and Engineering Program and KUIS AI Center, Koc University},
            city={Istanbul},
            postcode={34450}, 
            country={Turkey}}

\affiliation[inst2]{organization={Department of Mathematics, Koc University},
            city={Istanbul},
            postcode={34450}, 
            country={Turkey}}     

\affiliation[inst3]{organization={Department of Anatomy, School of Medicine, Koc University},
            city={Istanbul},
            postcode={34450}, 
            country={Turkey}}       

\affiliation[inst4]{organization={Department of Radiology, School of Medicine, Haseki SUAM, Medical Sciences University},
            city={Istanbul},
            postcode={34265}, 
            country={Turkey}}              

\affiliation[inst5]{organization={Department of Computer Engineering, School of Medicine, and KUIS AI Center, Koc University},
            city={Istanbul},
            postcode={34450}, 
            country={Turkey}}  
            
\cortext[cor1]{Corresponding author}

\begin{abstract}
Segmentation networks are not explicitly imposed to learn global invariants of an image, such as the shape of an object and the geometry between multiple objects, when they are trained with a standard loss function. On the other hand, especially when there exist a limited amount of training data, incorporating such invariants into network training may help regularize the training provided that these invariants are the intrinsic characteristics of the objects to be segmented. This paper addresses this issue by introducing a new topology-aware loss function that penalizes topology dissimilarities between the ground truth and prediction through persistent homology. Different from the previously suggested segmentation network designs, which apply the threshold filtration on a likelihood function of the prediction map and the Betti numbers of the ground truth, this paper proposes to apply the Vietoris-Rips filtration to obtain persistence diagrams of both ground truth and prediction maps and calculate the dissimilarity with the Wasserstein distance between the corresponding persistence diagrams. The use of this filtration has advantage of modeling shape and geometry at the same time, which may not happen when the threshold filtration is applied. Our experiments on 4327 CT images of 24 subjects reveal that the proposed topology-aware loss function leads to better results than its counterparts, indicating the effectiveness of this use. 


\end{abstract}

\begin{keyword}
Topology, persistent homology, Vietoris-Rips filtration, encoder-decoder networks, aorta and great vessel segmentation, computed tomography.
\end{keyword}

\end{frontmatter}

\section{Introduction}
\label{sec:introduction}
Encoder-decoder networks have achieved state-of-the-art results for various segmentation problems. The training of these networks relies on minimizing a loss function, e.g., mean squared error and cross-entropy, which typically defines the loss of each pixel separately and aggregates these pixel-wise losses. This aggregation might be unweighted, assigning the unit weight to each pixel's loss, or weighted, giving higher loss weights to hard-to-learn pixels. In the latter case, the pixels' weights can be assigned beforehand and remains the same during training, e.g., giving higher weights for pixels close to object boundaries~\cite{ronneberger15} or belonging to the minority foreground classes~\cite{sudre17}. Alternatively, these weights can be adaptively changed during the training by modulating them based on the network performance, e.g., reducing the weights of easy-to-learn pixels for which the network gives high posteriors at a given epoch~\cite{lin17},~\cite{gunesli20}. 

\begin{figure}
\centering
\includegraphics[height=7.5cm]{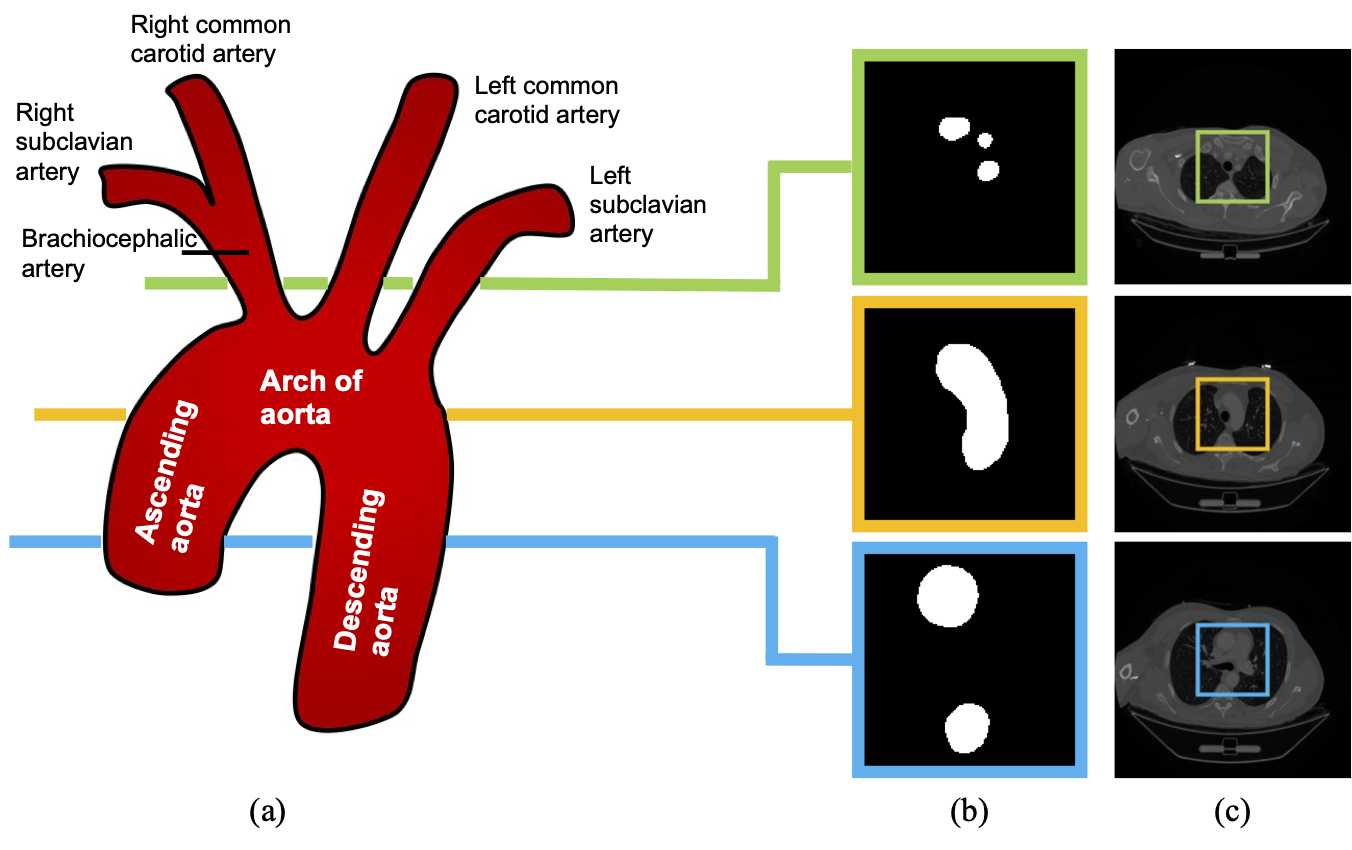} \vspace{-0.3cm}
\caption{(a) Anatomic formation of the aortic arch and the large arteries. (b) Manual annotations of the aortic arch and the large arteries on three exemplary axial slices. (c) CT scans for these annotations. Note that the annotations only for the green, orange, and blue squares are illustrated for better visualization. All pixels outside the rectangles are annotated as background.}
\label{fig:3d-aorta}
\end{figure}

These typical loss functions define the loss of each pixel only on its true and predicted values, but not considering those of other pixels, and aggregate them by weighted averaging or summing without considering the spatial relations between the predictions. Since this type of definition is of local nature, these loss functions may not sufficiently impose a network to learn the shape of an object or the geometry between multiple objects, especially when the amount of training data is small. On the other hand, the ability of the network to learn the shape may be important for better segmenting the objects in medical images since these objects typically have an expected shape or a geometry due to their intrinsic characteristics. One example is the formation of the aortic arch and great vessels in a human body. The aorta and the large arteries and veins (also known as great vessels) are not randomly distributed over the human body. Instead, they are found in a particular geometry due to the human anatomy (Figure~\ref{fig:3d-aorta}). Besides, they mostly seem as round objects on a 2D axial image since blood vessels are tubular in 3D. This anatomic information is indeed utilized by human annotators to locate these vessels and delineate their boundaries. 

\begin{figure}
\centering
\includegraphics[width=14.2cm]{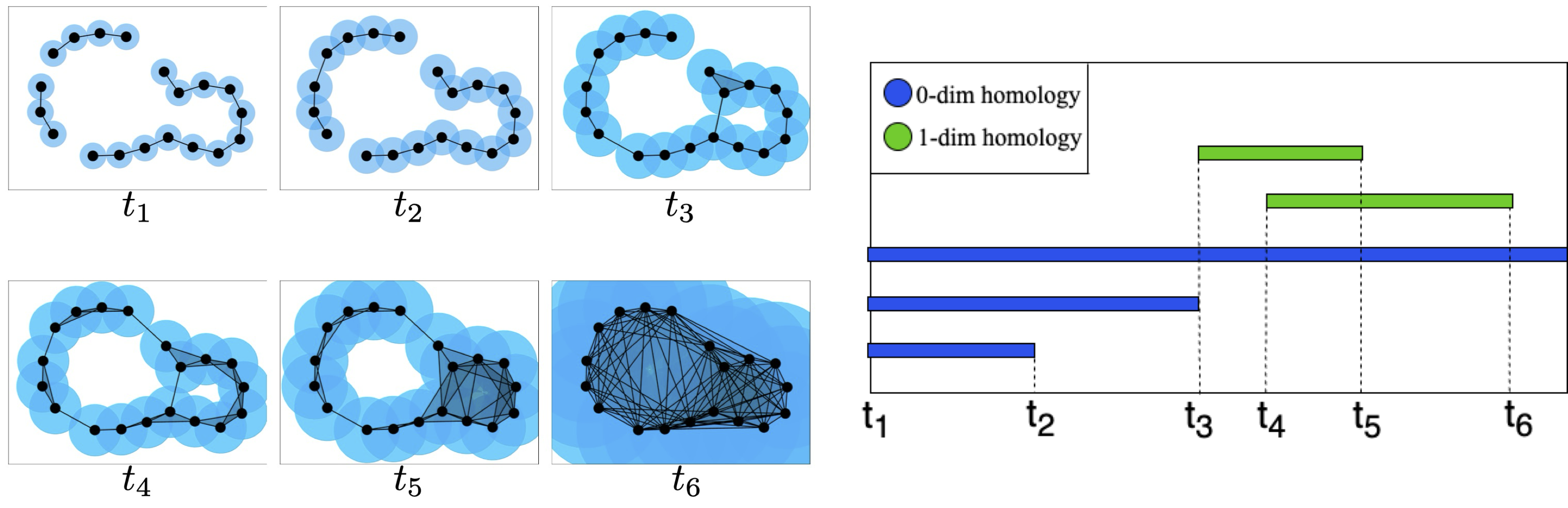}
\caption{Illustration of persistent homology. In this pedagogical example, we start with a point cloud consisting of the 19 black dots above. As time progresses, we gradually include more points in the filtration. At time $t,$ the associated level of the filtration consists of the blue colored region in diagram $t$ above. The rank of the 0-dimensional (resp. 1-dimensional) homology, which computes the number of connected components (resp. holes) in the filtration at time $t,$ is the number of intersections of the line $x=t$ with the set of blue lines (resp. green lines) in the last diagram. The homology groups at each level of the filtration can be computed via the corresponding Vietoris-Rips simplicial complexes as indicated above. The blue and green lines, which are called the barcodes for the persistent homology, give a summary of the rough shape of the configuration of points. The above times $t_1, \cdots, t_6$ are chosen to represent the precise levels where the ranks of the homology groups change, and hence are precisely the levels where the blue and green lines are born or die.}
\label{fig:ph-intro}
\end{figure}

In response to this issue, this paper introduces a new topology-aware loss function to train an encoder-decoder network for image segmentation. This loss function is defined as a weighted cross-entropy, in which the weight for a training sample (and thus, for its pixels) is calculated inversely proportional to topological similarity between the maps of its ground truth and predicted objects. This paper proposes to quantify topological features of these maps through persistent homology. Persistent homology is a mathematical tool that is a rough measure of the number of connected components, when computed in degree 0, and of the number of holes, when computed in degree 1 (Figure~\ref{fig:ph-intro}). Its main advantage compared to the more classical notion of homology is that, it is robust with respect to the addition of noise. We apply persistent homology by calculating the persistence diagram of each map associated to the Vietoris-Rips filtration on the point cloud of its object contours and enforce the network to minimize the Wasserstein distance between the corresponding persistence diagrams by defining the loss weight as a function of this distance. 

The proposed approach differs from the existing studies in the construction of the topological loss function and its integration to the network. Although there exist recent studies, such as~\cite{neur} and~\cite{pami}, that satisfactorily use persistent homology to train neural networks using the {\it topology} of the ground truth, our approach is different from these studies. We use persistent homology to learn {\it not only the topology but also the geometry} of the ground truth. Here, by geometry we mean the differential geometric nature of the objects in question, in the dataset we use namely the shape of the aorta and the distribution of the great vessels with respect to each other.

We will describe the details of this contribution in more mathematical terms below in Section~2. In summary, the difference of our contribution from the other studies stems from the type of filtration that we use in persistent homology, from the employment of the full strength of the persistent homology on the ground truth, as well as the use of the special metric to compare the persistent diagrams of the ground truth and prediction maps. The previous studies use the persistent homology of the prediction based on the threshold filtration associated to a likelihood function predicted by the network and the Betti numbers of the ground truth. Additionally, the loss functions they use only ensure that the \textit{topologies} of the ground truth and the prediction approached each other. On the other hand, different from this previous approach, in this paper, we propose to use the Vietoris-Rips filtration on the persistent homology of both the ground truth and the prediction maps and the Wasserstein distance between the corresponding persistence diagrams. The use of the Vietoris-Rips filtration takes into account the \textit{geometry} of the ground truth and the prediction maps, and the loss function based on the Wasserstein distance ensures that these two \textit{geometries} approach each other.

Besides, this is the first proposal of using a topology-aware loss function in a neural network design for the purpose of segmenting the aortic arch and great vessels, which are indeed found in a particular geometry in the human body, and thus, provides an exemplary showcase to demonstrate the usefulness of the geometry-preserving property of a neural network. Aortic arch and great vessel segmentation is important since it is often the first step in quantifying their characteristics and understanding their variations of clinical relevance. For example, existing vessel variations might cause complications during insertion of carotid artery stents or anterior decompression of the cervical spine. Although there exist previous network designs to segment the aorta and coronary arteries~\cite{cheung21},~\cite{gu21}, none of them use persistent homology in their models or in the definition of their loss functions.

\section{Related Work and Contribution}
\label{sec:related-work}

\subsection{Persistent Homology} 
For visual data of a medical nature, there are prior restrictions on their shape coming from the human anatomy. At first sight, the use of topological invariants like the rank of homology groups (known as Betti numbers) seems reasonable; however, these invariants are rigid, and summarizing noisy data will lead to complications. A solution to this problem is to use persistent homology, which is more stable for noisy data and is briefly reviewed in our context. 

An important  invariant of a topological space $X$ is the $n$-th homology group ${\rm H}_{n}(X),$ for each $n\geq 0$~\cite{hatcher02}. Here and elsewhere in this paper, we always consider homology groups with coefficients in the field of rational numbers $\mathbb{Q}.$ This makes the homology groups vector spaces over $\mathbb{Q}.$ The dimension $b_{n}(X)$ of the vector space ${\rm H}_{n}(X)$ is the $n$-th Betti number of $X$ and is a measure of the number of $n$ dimensional spheres which do not bound an $n+1$ dimensional ball. Thus, $b_{0}(X)$ is the number of connected components of $X$, and $b_{1}(X)$ is the number of circles that do not bound a disk. Even though the Betti invariants are extremely useful in the abstract study of topological spaces, they are too rigid to be of use in machine learning applications.

A variant of homology that is more convenient for real-world applications is the persistent homology~\cite{carlsson09},~\cite{chazal21}. Here the input is a topological space $\mathbb{X},$ which is endowed with a filtration $\{X_{t}\}_{t \in \mathbb{R}},$ indexed by the real numbers $\mathbb{R},$ with $\mathbb{X}=\cup _{t \in \mathbb{R}}X_{t}.$ The condition for being a filtration is that for every $s\leq t,$ $X_s \subseteq X_t.$ Taking homology of the spaces in the filtration for a fixed integer $n,$ we obtain a persistence module $\{ {\rm H}_{n}(X_{t}) \}_{t \in \mathbb{R}},$ which associates a $\mathbb{Q}$-vector space to each $t \in \mathbb{R}$ and a linear map between these vector spaces for each pair $(t,s)$ with $t\leq s,$ coming from the functoriality of homology. Associated to a persistence module, there is a barcode and a persistence diagram that summarizes at which filtration index the holes are born and at which filtration index they die. The flavor of the persistent homology and what it measures depends very much on which filtration one chooses to consider on $\mathbb{X}.$ Figure~\ref{fig:ph-intro} illustrates the generation of a barcode from a point cloud when the Vietoris-Rips filtration is used.

\subsection{Persistent Homology for Segmentation Networks}

There exist only a few studies that employ persistent homology in the design of a segmentation network. Similar to ours, this is achieved through the definition of a topological loss function. On the other hand, the main difference between these previous studies and ours is the type of filtration, the choice of which affects the phenomena persistent homology quantifies, and hence, the phenomena that a network is forced to learn during its training. In the context of segmentation networks, there are essentially two different ways one obtains a topological space with a filtration. 

One of these filtrations, which is the one that we employ in this work, is through the use of a distance function. Here, one starts with a point cloud $X$ in a metric space $M$ with a distance function $d.$ For each $t\in \mathbb{R},$ $X_{t}$
is the set of $m\in M$ such that there exists an $x \in X$ with $d(m,x)\leq t.$ Then $\{ X_{t}\}_{t \in \mathbb{R}}$ gives a filtration of $M.$ The persistence homology of this filtration encodes information about the shape of $X.$ This filtration is called the distance filtration or the Vietoris-Rips filtration below. The other type of filtration is constructed by using a real-valued function $f$ on space. If one lets $X_{t}:=f^{-1}((-\infty, t])$ then   $\{ X_{t}\}_{t\in \mathbb{R}}$  forms a filtration of the underlying space. In most of the studies below, $f$ is chosen to be $1-p$ where $p$ is a likelihood function on $\mathbb{R}^n,$ which aims to predict a shape $X$ in $\mathbb{R}^n.$ More precisely, we wish  $p$  to have the property that $x \in X$ if and only if $p(x)=1.$

\textit{The method of \cite{neur}}: In this work, if $\Omega$ denotes the image, which is viewed as a rectangular domain, there are two filtrations obtained on $\Omega,$ which correspond to two different functions on $\Omega.$ The first one is a binary function $f$ which assumes the value 0 on the foreground and 1 on the background. The other function is $g:=1-p,$ where $p$ is the likelihood function predicted by the neural network. The functions $f$ and $g$ give two different filtrations on $\Omega$ and these result in two different persistent homology data. The topological part of the loss function used in \cite{neur} is the square of the 2-Wasserstein distance between the persistence diagrams for these filtrations for both dimensions 0 and 1. The effect of using this topological loss function, in addition to the per-pixel cross-entropy loss, is that the network will emphasize learning the $0$-th and $1$-st Betti numbers of the ground truth, in addition to learning the pixels.

\textit{The method of \cite{pami}}: In this work, the topology of the ground truth is a priory knowledge. The topological loss function is constructed in terms of Betti numbers to force the network to make predictions that preserve the same topological structure as the ground truth. The loss function is based on increasing the barcode length of the $k$-th largest barcode lengths if $k$ is the desired Betti number. The same authors extended this method to multi-class image segmentation, including the Betti numbers corresponding to the triplets of the objects of different classes into the prior topology~\cite{itmi}. The method of \cite{haft20} is somewhat similar. In this paper, the authors define filtration by using the voxel intensity function on the data. The topological loss is the 1-Wasserstein distance between the persistence diagram of the prediction and the persistence diagram of the expected topological space.

\textit{The method of \cite{wong21}}: In this work, the authors use persistent homology in two different ways to improve the 3D segmentation of objects: First, they use a topological loss function in a similar vein as those in~\cite{pami} and~\cite{haft20} based on the likelihood function. Besides, the authors integrate persistent homology with a graph convolution network to capture multi-scale structural information. To do so, they form a point cloud in three dimensions and calculate persistence diagrams in each dimension using the distance filtration. Even though they incorporate the same filtration, this use involves adding the vectorization of the persistence image as a feature, which differs considerably from our method. Both of the uses in~\cite{wong21} do not define any loss function to minimize.

\textit{The method of \cite{topAwareFocal}}: 
In this study, the authors define filtration based on grayscale voxel intensities, where the filtration indices are discrete values ranging from zero to 255.  They generate persistence diagrams of 0, 1, and 2-dimensional holes from the voxel spaces based on this filtration. The Wasserstein distance between these diagrams is then computed using an optimal transportation plan derived from the Sinkhorn-Knopp algorithm. The objective is to minimize this distance in conjunction with the focal loss of the network's predictions.

\textit{Other uses}: In \cite{wang20}, the authors use persistent homology for a generative adversarial network to synthesize more realistic images in its generator. They map synthetic and real images into a topological feature space and define their topological dissimilarity as an additional loss term. In~\cite{mosinska18}, the authors define a topological loss term for a segmentation network, but not using persistent homology. Instead, they measure the difference between the pretrained VGG19 responses of the predicted and ground truth maps and use it as an additional loss term to correct the topology of linear structures in the ground truth. In~\cite{clDice}, the authors aim to address disconnectivity, and consequently broken topology, in the segmentation of thin foreground structures. They achieve this by formulating a loss function that focuses on the centerlines of thin, elongated foreground pixels by assigning greater weight to them.

\subsection{Persistent Homology for Other Network Tasks}

Classification networks use persistent homology as a tool to obtain better latent representations, which reflect the topological characteristics in the data. Such representations are obtained from persistence landscape~\cite{icml} and using the filtration associated to the height function~\cite{hofer17}, and integrated as a topological layer of the classification network. There are works using Vietoris-Rips filtration. In~\cite{hofer19}, a loss function is defined based on the death times of the barcodes for 0-dimensional persistent homology of the latent representation. Likewise, in~\cite{moor20}, connectivity properties of the latent representation in an autoencoder are improved through the use of a loss function based on the 0-dimensional persistent homology. Other principal uses of persistent homology in machine learning, which will not have relevance for this work, includes regularizing the weights of a network~\cite{gabrielsson20}, interpreting the weights of layers in a convolutional neural network~\cite{gabrielsson18}, and extracting topological features for a classifier~\cite{qaiser19}. Nevertheless, none of these studies use persistent homology to define a loss function for a segmentation network.

\section{Methodology}

Our method relies on 1) quantifying the topological features of the ground truth and the predicted segmentation maps by their persistence diagrams, 2) defining a loss function using the Wasserstein distance between the persistence diagrams of the ground truth and the prediction, and 3) training an encoder-decoder network by minimizing the proposed loss function. The following subsections give the details. 

\subsection{Persistence Diagram Calculation for the Aortic Arch and Great Vessels}\label{subsection-our-methodology}
\label{sec:pd}

Suppose that we start with a point cloud $X$ in $\mathbb{R}^n.$ In our case $n=2$, and $X$ will be the contours of the aorta and the great vessels for either the ground truth or the prediction of the network at a given epoch. For $t \in \mathbb{R},$ if we let $X_{t}$ to be the set of points in $\mathbb{R}^n,$ whose distance to $X$ is less than or equal to ${\rm max}(t,0),$ then this gives us a filtration $\{ X_t\}_{t \in \mathbb{R}}$ of $\mathbb{R}^n=\mathbb{X},$ with $X_{0}=X.$ The associated persistence diagrams can be thought of as more dynamic and stable versions of the ordinary homology groups of $X.$ 
\begin{figure}
\centering
\small{
\begin{tabular}{@{~}c@{~}cc@{~}c@{~}}
%
\includegraphics[width=2.8cm]{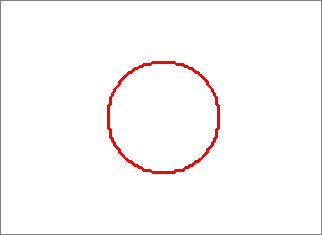} &
\includegraphics[height=1.8cm]{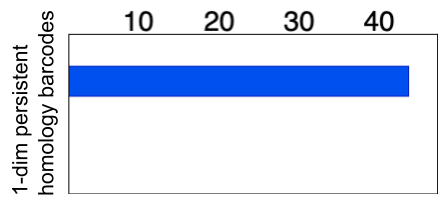} &
\includegraphics[width=2.8cm]{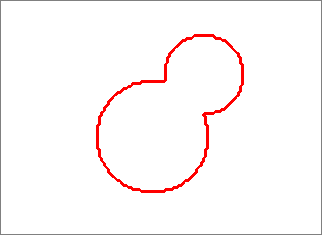} &
\includegraphics[height=1.8cm]{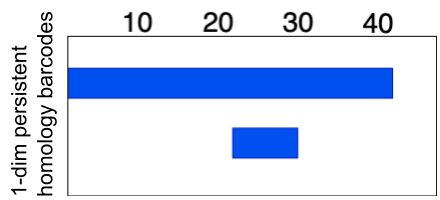} \vspace{-0.3cm} \\
(a) & (b) & (c) & (d) \\
\end{tabular}
}
\caption{(a), (c) Boundaries of two homotopy equivalent objects, and (b), (d) 1-dimensional persistent homologies of the objects shown in (a) and (c), respectively. Even though the standard homology groups of (a) and (c) are the same, since they are homotopy equivalent, their persistent homologies are not. The latter takes into account the shape of the object in this case. Namely, the smaller circular bump in the upper right part of (c) is responsible for the smaller bar in (d).} 
\label{fig:homotopy-equivalent-big}
\end{figure}

\begin{figure}
\centering
\small{
\begin{tabular}{@{~}c@{~}cc@{~}c@{~}}
\includegraphics[width=3.0cm]{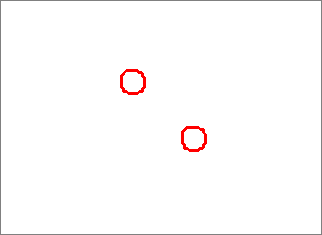} &
\includegraphics[width=4.6cm]{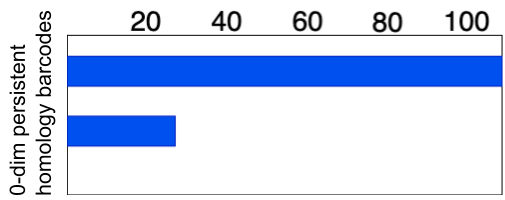} &
\includegraphics[width=3.0cm]{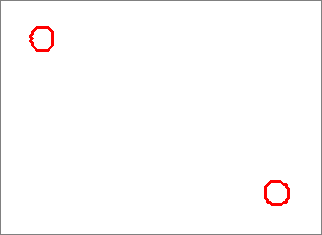} &
\includegraphics[width=4.6cm]{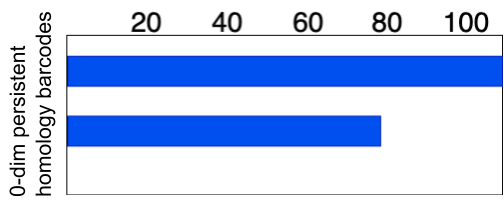} \vspace{-0.3cm} \\
(a) & (b) & (c) & (d) \\
\end{tabular}
}
\caption{(a), (c) Boundaries of two pairs of homotopy equivalent objects, and (b), (d) 0-dimensional persistent homologies of the objects shown in (a) and (c), represented by two bars. The long bar in each case represents the connected component which persists forever. The smaller bar in each case corresponds to the distance between the two connected components. This bar is smaller in (b) than in (d), consistent with the fact that the components are closer to each other in (a).} 
\label{fig:homotopy-equivalent-small}
\end{figure}

The persistent homology with the Vietoris-Rips (distance) filtration above has an added, somewhat surprising, benefit. It tells us about the {\it geometry} of $X,$ and not only about its topology. By geometry, we mean the shape of an object and also the distribution of multiple objects with respect to each other. Figure~\ref{fig:homotopy-equivalent-big} depicts the boundaries of two homotopy equivalent objects of different shapes, exhibiting the same topology but different 1-dimensional persistent homologies. Likewise, Figure~\ref{fig:homotopy-equivalent-small} sketches the boundaries of two pairs of homotopy equivalent objects with different distributions. These object pairs have the same topology, but this time, different 0-dimensional persistent homologies. Such geometry differences cannot be captured by a filtration associated to the likelihood function, as suggested by the previous segmentation networks~\cite{pami,neur,wong21}. On the other hand, as illustrated in these figures, the Vietoris-Rips filtration that we use in our design produces different barcodes, which allows us to model differences in the object geometries.

This idea will be very important for our segmentation model since the shape of the aorta and the distribution and the distances of the great vessels with respect to each other give us essential global invariants, which will help us improve the network using this prior geometric information. In our model, we define a loss function based on the 0-dimensional persistent homology if the ground truth includes any great vessels, which are indeed smaller in size compared to the aorta. The reason is that when we are dealing with the great vessels, the geometry of the associated point cloud is essentially determined by the distribution and the distances of the connected components in the data. The connected components in the images we consider correspond to the individual great vessels themselves. Even though the number of the connected components, hence the 0-th Betti number, in two point clouds might be the same, the corresponding barcodes associated to their 0-dimensional persistent homology might be quite different (see Figure~\ref{fig:homotopy-equivalent-small}). If the ground truth includes only the aortic arches, we use the 1-dimensional persistent homology in the loss function since this time the shape becomes more distinctive for these relatively larger veins. The fact that the holes are born and they die at different indices of the Vietoris-Rips filtration gives essential information about the shape of the aortic arch (see Figure~\ref{fig:homotopy-equivalent-big}), and this trait can be successfully used to train the network.

\subsection{Topology-Aware Loss Function}

\begin{figure*}[t]
\centering
\small{
\begin{tabular}{@{~}c@{~}}
\includegraphics[height=3.32cm]{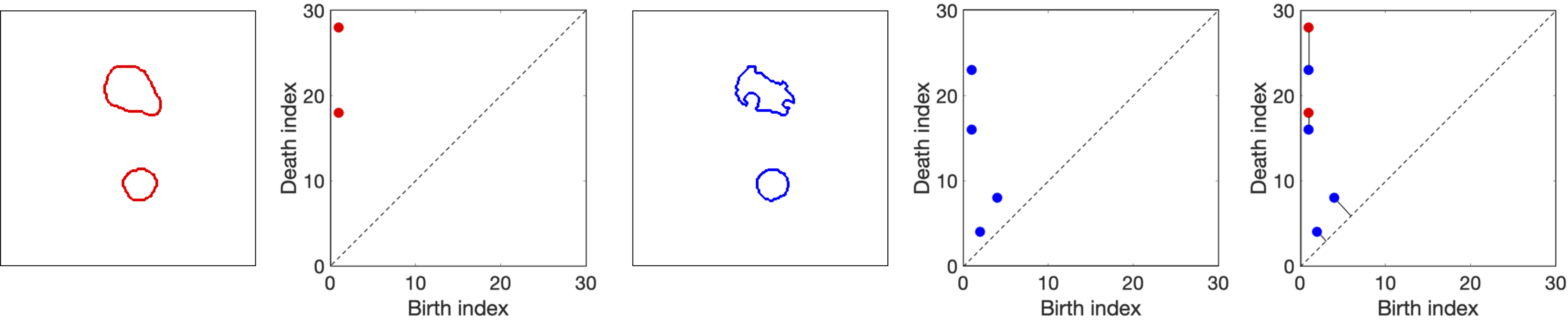} \\
(a)
\hspace{2.95cm}(b)
\hspace{2.75cm}(c)
\hspace{2.95cm}(d)
\hspace{3.0cm}(e)\\
\end{tabular}
}
\caption{(a) Boundaries of the objects found in the ground truth map ${\cal S}_I$ and (b) its persistence diagram $\Pi_{{\cal S}_I}$ for the homology group 1. (c) Boundaries of the objects found in the prediction map ${\widehat{\cal S}}_I$ and (d) its persistence diagram $\Pi_{{\widehat{\cal S}}_I}$ for the homology group 1. (e) Illustration of the best matching of these two diagrams. The Wasserstein distance is the sum of the powers of the distances between the matched points.} 
\label{fig:wasserstein}
\end{figure*}

The proposed topology-aware loss function $TopLoss= \sum_{I} \omega_I~L_I$ can be defined as a weighted sum of an arbitrary base loss function $L_I$, where the topological weight term $\omega_I$ for the training image $I$ calculated based on the difference between the persistence diagrams of its ground truth map ${\cal S}_I$ and the prediction map ${\widehat{\cal S}}_I$ estimated by the network at the end of each forward pass; these persistence diagrams are denoted by $\Pi_{{\cal S}_I}$ and $\Pi_{{\widehat{\cal S}}_I}$, respectively. We define the topological weight term $\omega_I$ as a linear combination of the Wasserstein distances of the homology group 0 and the homology group 1, $d_0(\Pi_{{\cal S}_I}, \Pi_{{\widehat{\cal S}}_I})$ and $d_1(\Pi_{{\cal S}_I}, \Pi_{{\widehat{\cal S}}_I})$, respectively, 
\begin{equation}
\omega_I = 1 + \alpha_I \cdot d_0(\Pi_{{\cal S}_I}, \Pi_{{\widehat{\cal S}}_I}) + \beta_I \cdot d_1(\Pi_{{\cal S}_I}, \Pi_{{\widehat{\cal S}}_I})
\label{eqn:term}
\end{equation} 
where $\alpha_I$ and $\beta_I$ are the constants that determine the importance of a homology group. Based on our discussions given at the end of Section~\ref{sec:pd}, if the ground truth ${\cal S}_I$ of the image $I$ includes any great vessel, we consider only the homology group 0 and empirically set $(\alpha_I$, $\beta_I)$ = (5.0$e$\textendash6, 0.0). Otherwise, if ${\cal S}_I$ contains only the aortic arches (without any great vessel), we consider the homology group 1 and set $(\alpha_I$, $\beta_I)$ = (0.0, 1.0$e$\textendash4). The calculation of the distance between two persistence diagrams relies on finding matches between the points in these two persistence diagrams that minimize the cost over all matchings, where the points are allowed to be matched with any point on a diagonal (Figure~\ref{fig:wasserstein}). The Wasserstein distance is defined as the sum of the powers of the distances between the matched points.

\subsection{Network Training}

\begin{figure}[t]
\includegraphics[height=8.4cm]{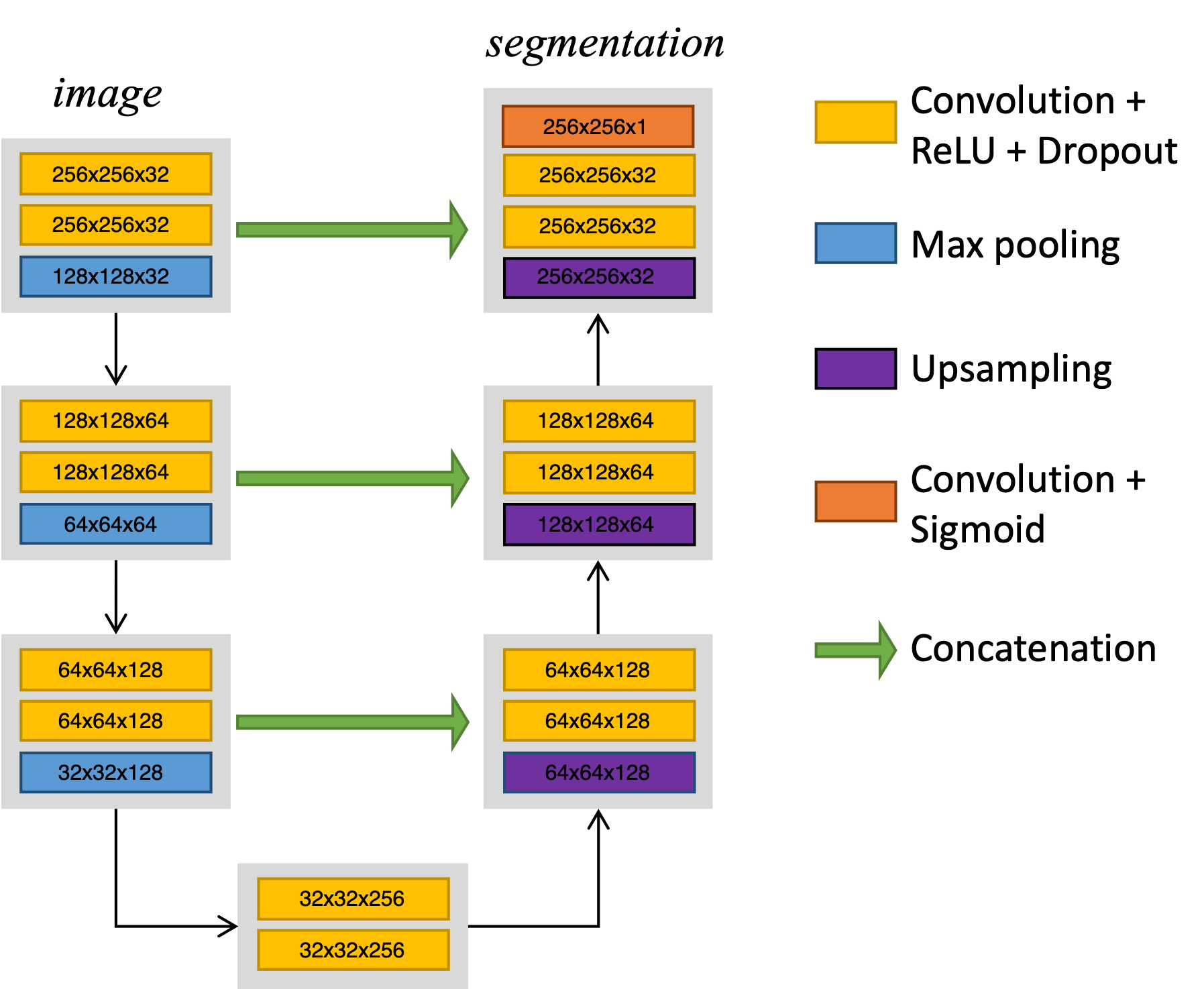}
\centering
\caption{UNet architecture used as the base model. Each box represents an operator. The first two numbers inside a box denote the height and the width of the inputs that the corresponding operator takes, and the last one is the number of feature maps used by this operator, respectively. In this network, all convolutions use a $3\times3$ filter, while pooling/upsampling operations use a $2\times2$ filter. The dropout factor is selected as 0.3.}
\label{fig:unetModel}
\end{figure}

We use an encoder-decoder network with the UNet architecture illustrated in Figure~\ref{fig:unetModel}. It is trained to minimize the proposed topology-aware loss function by backpropagation. In this study, we first chose the cross-entropy loss $\text{CE}_I$ as the base loss function $L_I$, and defined our topology-aware loss function as $TopLoss = \sum_{I} \omega_I~\text{CE}_I$. We then explored the effects of selecting other base loss functions in the experiments (Section~\ref{sec-loss-comparisons}). At each epoch of backpropagation, the forward pass estimates segmentation maps for every training image $I$ and updates the topology-aware loss $TopLoss$ by calculating cross-entropy losses $\text{CE}_I$ as well as topological weight terms $\omega_I$ with respect to the difference between the ground truths and the predictions. Then, the backward pass updates the network weights by differentiating the updated $TopLoss$. This training has a warm-up period for 25 epochs where only the base loss is used (i.e., $\omega_I = 1$). Afterwards, it continues with minimizing the proposed topology-aware loss function $TopLoss$. It is worth noting that this strategy is also used by the previous studies~\cite{neur}.

This network was implemented in Python using the PyTorch framework. It is available at https://github.com/seherozcelik/TopologyAware. In this implementation, persistence diagrams were computed using Ripser++~\cite{zhang2020gpu}, a GPU-accelerated topological data analysis tool, which utilized a C++ library as its core computational engine. Distance calculations were performed using the Gudhi library (https://gudhi.inria.fr), a topological data analysis framework also implemented in C++ with a Python interface. In the experiments, we end-to-end trained the network from scratch with an early stopping approach; training was stopped if there was no improvement on the validation set loss in the last 40 epochs. The AdaDelta optimizer with an initial learning rate of 1.0 was used to adaptively adjust the learning rate and the momentum. The batch size was selected as 1. The training was conducted on a Tesla T4 GPU.

\section{Experiments}
\subsection{Dataset}

The proposed topology-aware loss function was tested on a dataset that contains CT scans of 24 subjects with prediagnosis of pulmonary embolism. The CT scans were acquired using a 128 slice Philips Ingenuity CT scanner with 1.5 mm slice thickness. A 60 ml of non-ionic contrast material (iohexol; generic name Opaxol) was introduced with a 100 ml saline chaser at 5 ml/s. The data collection was conducted in accordance with the tenets of the Declaration of Helsinki and was approved by Koc University Institutional Review Board (Protocol number: 2022.161.IRB1.064). We randomly split the 24 subjects into the training and test sets. The training set contains 2896 images of 16 subjects; 2234 images of 12 subjects were used to learn the network weights by backpropagation and 662 images of 4 subjects were used as validation images for early stopping. The test set comprises 1431 images of 8 subjects; note that the images of none of these subjects were used neither in the training nor for early stopping.

\subsection{Evaluation}

Predictions were quantitatively evaluated by calculating the performance metrics both at the pixel- and vessel-level. For pixel-level evaluation, true positive pixels were found, and the precision, recall, and f-score were calculated for each image, separately. These metrics were then averaged over the test set images. For vessel-level evaluation, each vessel (a great vessel or an aortic arch) in the ground truth was matched with its maximally overlapping object in the prediction map, and considered as true positive if the intersection-over-union for this match was greater than 50 percent. Afterwards, true positive vessels were accumulated over all test set images and the vessel-level precision, recall, and f-score metrics were calculated. Additionally, the Hausdorff distance was calculated between each ground truth vessel and its maximally overlapping object in the prediction map, and vice versa. If there is no overlap for a vessel, the Hausdorff distance was calculated between this vessel and the closest segmented object. Then, for a test image, the overall Hausdorff distance was the weighted average of all Hausdorff distances where the weight of a vessel was selected as the ratio of the vessel's area to the area of all vessels in the ground truth. Better segmentations yield higher precision, recall, and f-score metrics, and lower Hausdorff distances.

\subsection{Comparisons and Ablation Studies}

We used five algorithms for comparison and ablation studies.
The first one was the \textit{Baseline} algorithm that used the same UNet architecture (Figure~\ref{fig:unetModel}) and training settings as ours except that it used the standard cross-entropy as its loss function. We used this algorithm in our comparisons to understand the importance of using a topology-aware loss function in the network training. Note that we used the same set of initial network weights for this baseline and our model to directly observe the effects of a loss function in a controlled setting.

The second algorithm had also the same network design, with the same set of initial network weights, but used another topology-preserving loss function suggested by~\cite{neur}. As mentioned in the introduction and the related work, this suggested loss relied on calculating the persistent homology based on the threshold filtration associated to a likelihood function predicted by the network and the Betti numbers of the ground truth. We included this \textit{LikelihoodFiltration} algorithm in our comparisons to investigate the benefits of using the Vietoris-Rips filtration on the persistent homology, which is effective in modeling the topology but also the geometry of the ground truth vessels (i.e., both the shape of the vessels and the distribution of the vessels with respect to each other). The third comparison was with the \textit{FourierNet} algorithm that proposed a shape-preserving network design~\cite{fourierNet}. This algorithm represented the shape prior by extracting Fourier descriptors on the objects' contours and concurrently learned these descriptors with the main task of segmentation. We included this algorithm in our comparisons to observe the effects of modeling the vessels' geometry instead of modeling only the vessels' shape. Note that in our experiments, we run these algorithms using the codes provided by its authors.

The last two algorithms were to demonstrate the adaptability of the proposed loss function framework to other base loss functions. For that, we first trained two networks using the standard Dice loss~\cite{milletari} and the \textit{BoundaryLoss} proposed in~\cite{boundaryLoss}. Then, we used each of these losses as the base loss of our topology-aware loss function, defining $TopLoss = \sum_{I} \omega_I~\text{Dice}_I$ and $TopLoss = \sum_{I} \omega_I~\text{BoundaryLoss}_I$, respectively, and compare our results with the corresponding baseline. 

\begin{table*}
\renewcommand{\arraystretch}{0.75}  
\caption{Performance metrics calculated over the test set images. In this table, our proposed topology-aware loss function $TopLoss = \sum_{I} \omega_I~\text{CE}_I$ used the cross-entropy $\text{CE}_I$ as its base loss. These are the averages and standard deviations across five runs.}\vspace{0.2cm}
\small{
\begin{tabular}{|@{~}l@{~}|c@{~}|@{~}c@{~}|@{~}c@{~}|c@{~}|@{~}c@{~}|@{~}c|c@{~}|}
\hline
& \multicolumn{3}{c|}{Pixel-level metrics} & \multicolumn{3}{c|}{Vessel-level metrics} &Hausdorff \\
\cline{2-7}
& Precision & Recall & F-score & Precision & Recall & F-score & distance \\
\hline
$TopLoss$ & \textbf{88.2$\pm$1.4} & \textbf{87.2$\pm$2.1} & \textbf{86.6$\pm$0.7} & 74.6$\pm$2.0 & 84.0$\pm$1.2 & \textbf{79.0$\pm$0.9} & \textbf{5.3$\pm$0.4}\\
Baseline with CE~\cite{ronneberger15} & 87.6$\pm$2.0 & 86.0$\pm$2.5 & 85.6$\pm$2.2 & 71.5$\pm$6.3 & 82.1$\pm$4.8 & 76.4$\pm$5.6 & 5.5$\pm$0.9\\
LikelihoodFiltration~\cite{neur} & 87.2$\pm$0.7 & 86.9$\pm$0.8 & 85.8$\pm$0.6 & 73.3$\pm$1.7 & \textbf{84.1$\pm$2.2} & 78.3$\pm$1.7 & 5.7$\pm$0.6\\
FourierNet~\cite{fourierNet} & 85.7$\pm$2.8 & 81.2$\pm$4.3 & 81.6$\pm$1.6 &  \textbf{76.9$\pm$3.1} & 73.1$\pm$2.9 & 74.9$\pm$1.5 & 7.6$\pm$1.1\\
\hline
\end{tabular}
}
\label{table:results}
\end{table*}

\subsection{Results and Discussion}

\begin{figure*}
\centering
\small{
\begin{tabular}{@{~}c@{~}c@{~}c@{~}c@{~}c@{~}c@{~}}
\includegraphics[width=2.50cm]{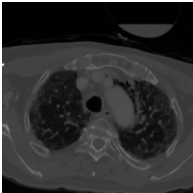} &
\includegraphics[width=2.50cm]{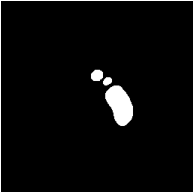} &
\includegraphics[width=2.50cm]{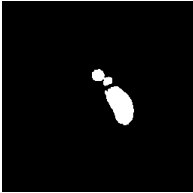} &
\includegraphics[width=2.50cm]{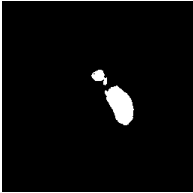}& \includegraphics[width=2.50cm]{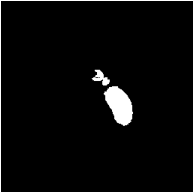} &
\includegraphics[width=2.50cm]{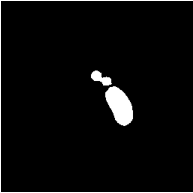} \vspace{-0.1cm}\\
\includegraphics[width=2.50cm]{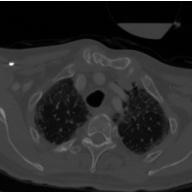} &
\includegraphics[width=2.50cm]{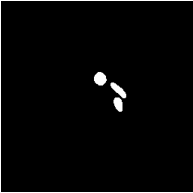} &
\includegraphics[width=2.50cm]{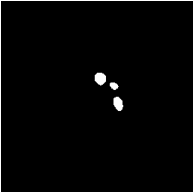} &
\includegraphics[width=2.50cm]{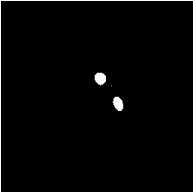} &
\includegraphics[width=2.50cm]{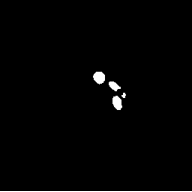} &
\includegraphics[width=2.50cm]{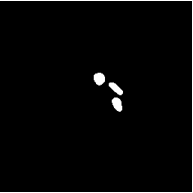} \vspace{-0.1cm}\\
\includegraphics[width=2.50cm]{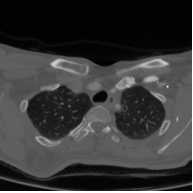} &
\includegraphics[width=2.50cm]{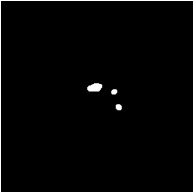} &
\includegraphics[width=2.50cm]{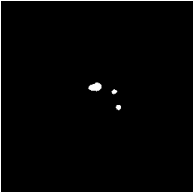} &
\includegraphics[width=2.50cm]{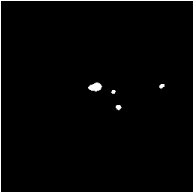} &
\includegraphics[width=2.50cm]{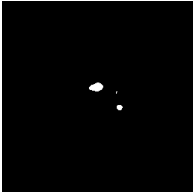} &
\includegraphics[width=2.50cm]{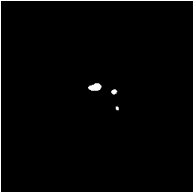} \vspace{-0.1cm}\\
\includegraphics[width=2.50cm]{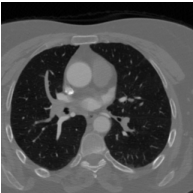} &
\includegraphics[width=2.50cm]{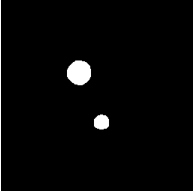} &
\includegraphics[width=2.50cm]{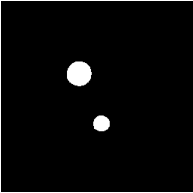} &
\includegraphics[width=2.50cm]{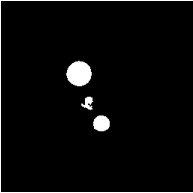} &
\includegraphics[width=2.50cm]{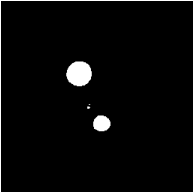} &
\includegraphics[width=2.50cm]{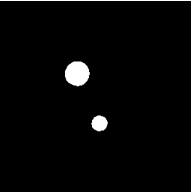} \vspace{-0.1cm}\\
\includegraphics[width=2.50cm]{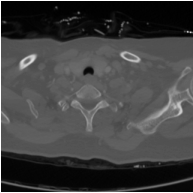} &
\includegraphics[width=2.50cm]{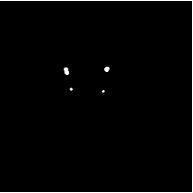} &
\includegraphics[width=2.50cm]{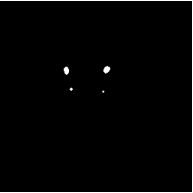} &
\includegraphics[width=2.50cm]{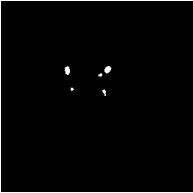} &
\includegraphics[width=2.50cm]{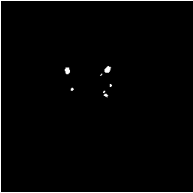} &
\includegraphics[width=2.50cm]{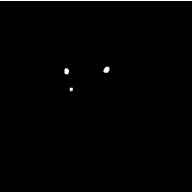} \vspace{-0.1cm}\\
\includegraphics[width=2.50cm]{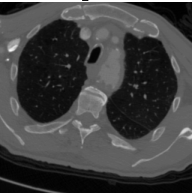} &
\includegraphics[width=2.50cm]{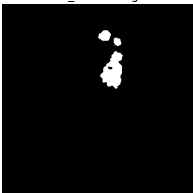} &
\includegraphics[width=2.50cm]{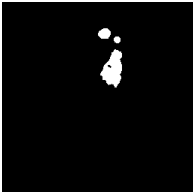} &
\includegraphics[width=2.50cm]{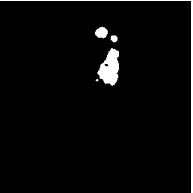} &
\includegraphics[width=2.50cm]{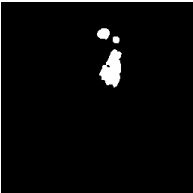} &
\includegraphics[width=2.50cm]{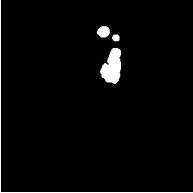} \vspace{-0.1cm}\\
\includegraphics[width=2.50cm]{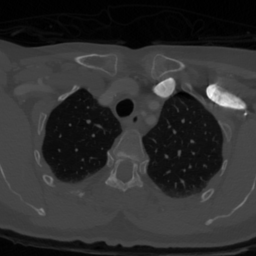} &
\includegraphics[width=2.50cm]{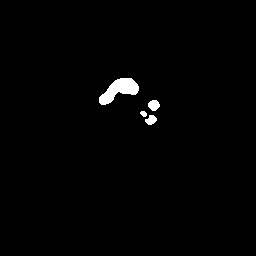} &
\includegraphics[width=2.50cm]{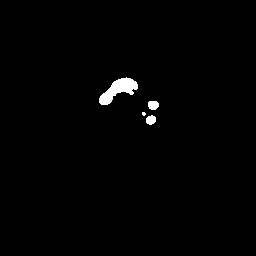} &
\includegraphics[width=2.50cm]{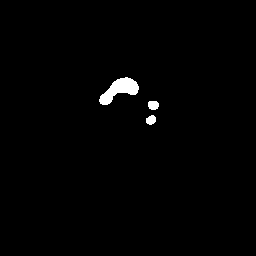} &
\includegraphics[width=2.50cm]{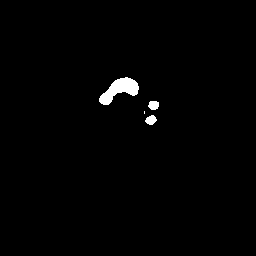} &
\includegraphics[width=2.50cm]{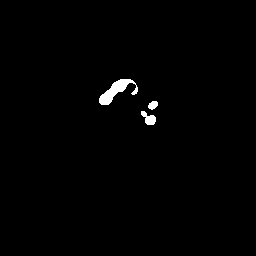} \vspace{-0.1cm}\\
(a) & (b) & (c) & (d) & (e) & (f)
\end{tabular}
}
\caption{(a) Example test set images. (b) Ground truths. (c) Results of the proposed topology-aware loss function $TopLoss = \sum_{I} \omega_I~\text{CE}_I$ with the cross-entropy $\text{CE}_I$ being used as its base loss. (d) Results of the baseline~\cite{ronneberger15} using the standard cross-entropy loss. (e) Results of the topology-preserving network, the LikelihoodFiltration algorithm, proposed by~\cite{neur}. (f) Results of the shape-preserving FourierNet algorithm proposed by~\cite{fourierNet}.} 
\label{fig:results}
\end{figure*}

Table~\ref{table:results} reports the test set results obtained by our topology-aware loss function $TopLoss$ that uses the cross-entropy as its base loss, and the first three comparison algorithms. We run all these models five times, and these are the quantitative results averaged across the five runs. This table reveals that our model led to better segmentations, giving higher f-scores and lower Hausdorff distances. The visual results obtained on exemplary test set images were also consistent with this observation (Figure~\ref{fig:results}). Comparing with the Baseline algorithm, which uses the standard cross-entropy as its loss function, our proposal was more effective to eliminate false negatives (first two rows of Figure~\ref{fig:results}) as well as to correct false positives (third to fifth rows of Figure~\ref{fig:results}). Although the \textit{LikelihoodFiltration} and \textit{FourierNet} algorithms can correct them to some extent, the proposed model was more effective than these algorithms, as also reflected in the quantitative results. Additionally, the example in the sixth row of Figure~\ref{fig:results} demonstrates that our proposed method captured the intricate details of a subtle shape more effectively than the other models. The last row of Figure~\ref{fig:results} presents an example of a successful correction of an undersegmentation. Despite the undersegmented vessel being very small and situated close to a neighboring vessel, the proposed method also effectively addressed this intricate challenge, whereas other methods failed.

\begin{table}
\renewcommand{\arraystretch}{0.75}  
\caption{Test set metrics calculated over the images containing any great vessels and only aortic arches. In this table, our proposed topology-aware loss function $TopLoss = \sum_{I} \omega_I~\text{CE}_I$ used the cross-entropy $\text{CE}_I$ as its base loss. These are the averages and standard deviations across five runs.}
\small{
\begin{tabular}{|@{~}l@{~}|c@{~}|@{~}c@{~}|@{~}c@{~}|c@{~}|@{~}c@{~}|@{~}c|c@{~}|}
\hline
\rowcolor[HTML]{cccccc} \multicolumn{8}{|l|}{Test images containing any great vessels} \\ \hline
& \multicolumn{3}{c|}{Pixel-level metrics} & \multicolumn{3}{c|}{Vessel-level metrics} &Hausdorff \\
\cline{2-7}
& Precision & Recall & F-score & Precision & Recall & F-score & distance \\
\hline
$TopLoss$ & \textbf{83.2$\pm$2.4} & \textbf{80.5$\pm$3.3} & \textbf{79.9$\pm$1.0} & \textbf{72.1$\pm$1.0} & 78.6$\pm$1.9 & \textbf{75.2$\pm$1.0} & \textbf{5.7$\pm$0.6}\\
Baseline with CE~\cite{ronneberger15} & 82.5$\pm$3.1 & 78.4$\pm$4.7 & 78.5$\pm$3.8 & 68.9$\pm$6.8 & 76.3$\pm$6.3 & 72.4$\pm$6.5 & 5.9$\pm$1.4\\
\textit{LikelihoodFiltration}~\cite{neur} &  81.8$\pm$1.2 & 80.0$\pm$1.5 & 78.8$\pm$0.9 & 71.4$\pm$1.9 & \textbf{79.2$\pm$3.1} & 75.1$\pm$2.1 & 6.0$\pm$0.6\\
\textit{FourierNet}~\cite{fourierNet} & 80.3$\pm$3.8 & 71.0$\pm$6.9 & 72.7$\pm$2.9 & 71.66$\pm$3.2 & 64.2$\pm$4.1 & 67.6$\pm$1.9 & 9.9$\pm$1.9\\
\hline
\rowcolor[HTML]{cccccc} \multicolumn{8}{|l|}{Test images containing only aortic arches} \\ \hline
& \multicolumn{3}{c|}{Pixel-level metrics} & \multicolumn{3}{c|}{Vessel-level metrics} &Hausdorff \\
\cline{2-7}
& Precision & Recall & F-score & Precision & Recall & F-score & distance \\
\hline
$TopLoss$ & \textbf{93.5$\pm$0.7} & \textbf{94.5$\pm$1.1} & \textbf{93.7$\pm$0.8} & 80.3$\pm$5.2 & \textbf{96.8$\pm$1.3} & 87.7$\pm$3.2 & \textbf{4.9$\pm$0.4}\\
Baseline with CE~\cite{ronneberger15} & 93.1$\pm$1.2 & 94.1$\pm$0.2 & 93.3$\pm$0.6 & 77.2$\pm$6.1 & 95.8$\pm$1.5 & 85.5$\pm$4.2 & 5.0$\pm$0.6\\
\textit{LikelihoodFiltration}~\cite{neur} & 93.2$\pm$0.9 & 94.2$\pm$0.9 & 93.3$\pm$0.7 & 77.5$\pm$3.2 & 95.9$\pm$1.6 & 85.7$\pm$2.4 & 5.4$\pm$0.8\\
\textit{FourierNet}~\cite{fourierNet} & 91.5$\pm$1.8 & 92.3$\pm$1.8 & 91.3$\pm$1.0 & \textbf{87.3$\pm$2.6} & 94.7$\pm$2.0 & \textbf{90.8$\pm$2.1} & 5.0$\pm$0.4\\
\hline
\end{tabular}
}
\label{table:results2}
\end{table}

The main contribution of this work is to use the Vietoris-Rips filtration for calculating the persistent homology of the ground truth and the prediction. This  has the benefit of modeling the shape of the objects and their {\it geometry}, which the persistent homology associated to a likelihood function fails to detect. This concurrent modeling is essential to capture the global invariants in our application. Since CT images contain both aorta and great vessels, the shape of the aorta and the distribution and the distances of the great vessels with respect to each other contain important prior geometric information that could be exploited. To investigate this further, we also calculated the performance metrics separately, for images containing any great vessels and for those containing only the aorta (or the aortic arches). These metrics are separately reported in Table~\ref{table:results2}. They demonstrate that the proposed topology-aware loss function improved the metrics both for the great vessels and the aortic arches. Note that the \textit{FourierNet} algorithm, with the shape-preserving property, gave the best vessel-level f-score for the aortic arches, which was consistent with our observation that the shape is important for the aorta. On the other hand, it was not successful to model the distribution of the great vessels, and in turn, it yielded the worst results for them.

\begin{table}[t]
\renewcommand{\arraystretch}{0.75}  
\caption{For different base loss functions, test set results obtained with and without using the proposed topology-aware loss function $TopLoss$ in the network training. These are the averages and standard deviations across five runs. Statistically significantly best metrics ($p < 0.05$) are indicated in bold.}
\small{
\begin{tabular}{|@{~}l@{~}|c@{~}|@{~}c@{~}|@{~}c@{~}|c@{~}|@{~}c@{~}|@{~}c|c@{~}|}
\hline
& \multicolumn{3}{c|}{Pixel-level metrics} & \multicolumn{3}{c|}{Vessel-level metrics} &Hausdorff \\
\cline{0-6}
Baselines & Precision & Recall & F-score & Precision & Recall & F-score & distance \\
\hline
Cross-entropy~\cite{ronneberger15} & 87.6$\pm$2.0 & 86.0$\pm$2.5 & 85.6$\pm$2.2 & 71.5$\pm$6.3 & 82.1$\pm$4.8 & 76.4$\pm$5.6 & 5.5$\pm$0.9\\
Dice loss~\cite{milletari} & 86.7$\pm$1.0 & 87.7$\pm$3.1 & 85.9$\pm$1.5 & 72.3$\pm$2.5 & 87.9$\pm$2.6 & 79.3$\pm$2.1 & 6.1$\pm$1.2\\
Boundary loss~\cite{boundaryLoss} & 88.0$\pm$1.2 & 87.2$\pm$1.2 & 86.1$\pm$0.5 & 78.9$\pm$1.7 & 85.1$\pm$1.3 & 81.8$\pm$1.1 & 5.0$\pm$0.3\\
\hline
\multicolumn{8}{|l|}{Proposed $TopLoss$}\\
\hline
Cross-entropy & \textbf{88.2$\pm$1.4} & \textbf{87.2$\pm$2.1} & \textbf{86.6$\pm$0.7} & \textbf{74.6$\pm$2.0]} & 84.0$\pm$1.2 & \textbf{79.0$\pm$0.9} & \textbf{5.3$\pm$0.4}\\
Dice loss & 86.8$\pm$0.5 & \textbf{88.6$\pm$1.0} & \textbf{86.6$\pm$0.7} & 74.0$\pm$3.7 & 88.4$\pm$1.1 & \textbf{80.5$\pm$2.5} & \textbf{5.4$\pm$0.5}\\
Boundary Loss & 87.9$\pm$0.7 & \textbf{88.4$\pm$0.9} & \textbf{86.7$\pm$0.4} & \textbf{81.5$\pm$1.2} & 85.9$\pm$1.1 & \textbf{83.6$\pm$0.6} & 4.9$\pm$1.4\\ \hline
\end{tabular}
}
\label{table:results3}
\end{table}

\subsection{Experiments with Different Losses}
\label{sec-loss-comparisons}

The motivation of defining the topology-aware loss function $TopLoss$ is to improve the training of a given network by explicitly enforcing it to preserve the topology of the segmented objects. This is achieved by introducing the topological weight term $\omega_I$ that penalizes the difference between the topology (persistence diagrams) of the ground truth and the prediction maps. The previous section reports the results when this term was incorporated into the cross-entropy loss $\text{CE}_I$. To explore the potential of incorporating the proposed framework into other losses, this section considers two additional loss functions, the standard Dice loss~\cite{milletari} and a custom loss called BoundaryLoss, proposed by another study~\cite{boundaryLoss}, and defines the topology-aware loss function as $TopLoss = \sum_{I} \omega_I~\text{Dice}_I$ and $TopLoss = \sum_{I} \omega_I~\text{BoundaryLoss}_I$, respectively. Table~\ref{table:results3} provides the comparison of our proposal and the baseline networks when these loss functions were used. One can have the following observations: First, as expected, a more complex base loss was more effective to better train the baseline network; the Dice loss and the boundary loss enhanced the f-scores of the network compared to the case when it was trained with the cross-entropy. On the other hand, defining the topology-aware loss function $TopLoss$ by incorporating the topological weight term $\omega_I$ into the corresponding base loss further improved the results of the corresponding baseline; these improvements were statistically significant with $p < 0.05$. These results demonstrated that the topological weight term $\omega_I$ can be integrated with various base losses and the proposed framework can be used to enhance the results of the corresponding baseline. 

\section{Conclusion}

This paper presented a topology-aware loss function for an encoder-decoder network. This was defined as a weighted loss function, in which the weight for an image was the topological dissimilarity between the ground truth map and the segmented map predicted by the network at the end of each epoch. Different from the previously suggested segmentation network designs, this paper proposed to apply the Vietoris-Rips filtration to obtain the persistence diagrams of these maps and calculate their (dis)similarity using the Wasserstein distance between the corresponding persistence diagrams. Experiments on 4327 CT images of 24 subjects revealed that this proposal was more effective than its counterparts in simultaneously modeling the shape of the aorta and the geometry between the great vessels. 

The dataset used in this study provided an exemplary showcase to demonstrate the effectiveness of the topology-aware loss since it necessitated modeling the shape of the aorta and the geometry between the great vessels at the same time. Thus, it was suitable for showing the effectiveness of applying the Vietoris-Rips filtration to obtain the persistence diagrams for defining the proposed topology-aware loss function. On the other hand, some other datasets, for example the ones focusing on segmenting one object (e.g., one organ in a CT image), may not require this simultaneous modeling, and thus, the use of just a shape-aware loss function may be sufficient. Additionally, the amount of training samples was kept as small since the use of such loss functions as a regularization technique in training is known to be more effective when the annotated data is limited, which is indeed very typical for various medical image segmentation tasks. In this work, we used the topological dissimilarity between the ground truth and the prediction to define the weight of an image in the loss function. In other words, we used the same dissimilarity metric to penalize every pixel in the same image, regardless of whether they were correctly or incorrectly predicted. One future research direction is to reflect the dissimilarity only to false negative and false positive pixels, and possibly with different extents.

\section*{Acknowledgments}

This work was partly supported by the Scientific and Technological Research Council of Turkey, project no T{\"U}B\.{I}TAK 120E497.

\section*{Declaration of generative AI and AI-assisted technologies in the writing process}

During the preparation of this work the authors used Google Translator, Grammarly, ChatGPT, and DeepL for editing and grammar enhancement. After using this tool/service, the authors reviewed and edited the content as needed and take full responsibility for the content of the publication.

\bibliographystyle{elsarticle-num} 
\bibliography{cas-refs}

\end{document}